# EFFECTIVE M-LEARNING DESIGN STRATEGIES FOR COMPUTER SCIENCE AND ENGINEERING COURSES

Ibrahim Alkore Alshalabi[1] and Khaled Elleithy[2]

[1]Department of Computer Science Engineering, University of Bridgeport,
Bridgeport- CT, USA
ialkorea@bridgeport.edu, Ibrkh75@yahoo.com

[2]Department of Computer Science Engineering, University of Bridgeport,
Bridgeport- CT, USA
elleithy@bridgeport.edu

## ABSTRACT

*Mobile learning (M-learning) is receiving more attention as a method of delivering to learners study materials anytime and anywhere. It is a necessity for educators to come up with a layout for learning that can be accessed through mobile devices. These learning materials should consist of good quality learning theories and accurate instructional layout in order to maintain the learning as effective as possible. It is important to follow certain strategies that can help the developers for M-learning applications. In this paper we proposed a set of strategies that are useful for creating mobile prototype for Computer Science and Engineering courses or M-learning application for course content.*

## 1. INTRODUCTION

One of the key criteria for any new technology to be successful is that it needs to be easy to learn and use. The new technology nowadays is mobile devices while the traditional medium for delivering E-learning is PC. To move from E-learning to M-learning may require dealing with some issues that will need to be taken into consideration during the construction of the content. Mobile learning is defined as the provision of education and training on mobile devices [1].
Mobile devices are likely to increase learning opportunities, the evolution and advancements of mobile devices will continue to accelerate as well. As a result, mobile devices will become a critical component in developing learning strategies. When developing learning strategies for the future, by applying one or more of the learning theories which can be used or implemented in our strategies, we will obtain better outcomes from our courses.

## 2. RELATED WORKS

Several studies have been done to strategies for M-learning and E-learning systems. Chang, Chen and Hsu [2], did a research that focuses on learners and resource recycle where possible and distinction through an educational website based on WebQuest. They illustrate the effect of different teaching strategies on the discovering efficiency of environment education and learning through quantitative methods. The individuals in this research were divided into three groups: conventional instruction, conventional WebQuest instruction, and WebQuest in distance education [2]. By using ANCOVA analysis, the outcomes show significant variations between the three categories in the post-examination. The research found two interesting outcomes. First, the process of using WebQuest in addition to real situations assisted students to acquire more experience and knowledge. Second, student knowledge immediately affected task efficiency and in a roundabout way made an effect on learning performance [2].

Just in Time Teaching (JiTT) features the characteristics of Blending Learning incorporated with the advantages of traditional Learning and E-Learning, applying a heart in teaching, creativity and tracking, embodying the effort, passion and entertainment in the success of learning as a student. More precisely, it is a teaching and learning strategy centered on the relationship between web-based study projects and an active student classroom [3]. Lei and Caixing have developed a new Instructional Design combined with M-learning and blending learning according to the instructional method of JiTT. The authors did the analysis on a software SuperMemo. New program can be used in system teaching which has been found, it is not only a program used to consider terms, but also discovering other understanding by segments of a topic by means of question-answer [3].

The reason of studying Mobile Learning & Commuting is to go more greatly into the regards between M-learning and commuting for a student account/profile that uses trips to execute university-learning activities. The research is to understand the needs and demands of students who study and/or do studies-related actions while commuting. This purpose includes disclosing the present behavioural styles of learners in mobile situations and knowing how the new gadgets and components that they could use, such as the e-book, sound content and mobile Internet, can assist in their study performance. An interview was done with learners from the Universitat Oberta de Catalunya (Open School of Catalonia, UOC)—a completely on-line college institution—in their travelling perspective on the train or bus. Next to that the observations on the attributes of the context were mentioned in [4].

## 3. LEARNING ENVIRONMENT

It's important to determine the components of the M-learning environment. The components M-learning systems are shown in Figure-1.

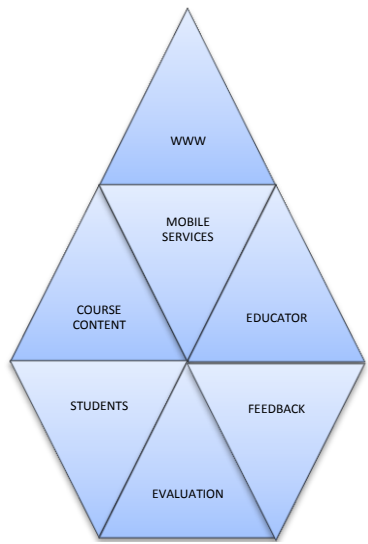

Figure-1: M-Learning environment component

## 3.1 Course content

To design or create course content, we need to analyze the course structure factors in terms of the main content, implementation and the goal of the course so that it can fit for use small devices with limited resources.

## 3.2 Students

Students are the main body of any teaching system. The purpose of our M-learning system is to teach the content using M-learning system recourse and enhancing the students' self-learning, knowledge and innovation capabilities.

## 3.3 Educator

Educators play a new role in M-learning system. The educators play the role of inspiring the innovative spirit of students and letting them embark initiatives. This approach gives students a chance to act according to their own information so they can achieve the self-feedback regarding to what they have learned. This is a new role added with the old role of the educators by controlling the course content, giving the order flow of the course content during the classes and organizing of the whole process [5].

## 3.4 Evaluation

For E-learning and M-learning environment we need an appropriate student's knowledge assessment. We need to use and apply observation on performance such as assessments, and offer information that is well equipped to help educators, students, and administrators.

## 3.5 Feedback

The feedback is the process of collecting information from all parts who are involved in the teaching process. This includes students, teachers, course developer, and evaluation and so on. The need of feedback is to find and fix drawbacks and the demerits to fix them and the benefits to emphasized and further improved. By using the feedback we can reach high level of quality education.

## 4. WHY IS M-LEARNING STYLE?

Different students have different modes of learning especially if attractive technology is involved. In order to better the students' educational abilities, the teaching approach is matched with the chosen learning style. Students believe that new technology mobile devices used for mobile learning can enhance the current learning practices in many aspects. This includes: When and where, better effective method of learning, reduce time to access learning materials, receive feedback, and act on time and so on.

## 5. COMPUTER SCIENCE AND ENGINEERING COURSES VS. LEARNERS

Computer engineering and science classes involve around practical courses and need extreme training, especially the programming course. [6]. Courses of science and engineering include theory and computer programming. These courses are important professional basic courses of computer science and electrical engineering students. For students to learn and master computer technology courses, content preparation is important to develop an application that covers the course contents. Students may find great difficulty in the M-learning application if they lack knowledge on the technology used. [7] [8]. Thus, content preparation is important in order to develop an application that covers the course content. The developer needs to know the features and interests of the learners by communicating with them, introducing to them the characteristic of the curriculum, and helping them with the course selection, the study plan, the time arrangement and progress control.

## 6. MODEL FOR DESIGN STRATEGIES COMPUTER SCIENCE AND ENGINEERING COURSES

The purpose of this paper is to develop strategies that can be followed during the creating of computer science or engineering course for M-learning environment.

### 6.1 Strategy 1

Computer science and engineering courses are different from other courses. In first strategy, these differences have to be considered during construction of the content. The many learning theories suggested need to be considered before construction of the content is readily adapted to teaching. During the creating of computer science or engineering course we need to use one or more of learning theories to work as a group does deliver high quality of knowledge to the learners.

The main theoretical bases for M-learning and E-learning are Behaviorism, Cognitivism, Constructivism, Humanism, Cooperative learning. By using these learning theories, depending on our course needs, we can get the following from each learning theory: Behaviorism regards that the individual's learning behavior is attributed to the individual's adaptation to the external environment. The cause of learning is considered the response to external stimuli; therefore, controlling stimuli can control the behavior and predict behavior, and thus can control and predict learning. Behaviorism does not care about the internal psychological process arising from external stimuli, regards that learning has nothing to do with internal psychological processes. Cognitivism learning theory is the reorganization of knowledge structures; learning is a relatively lasting change of ability or inclination, caused by the experience; learning process is the process of information processing. Constructivism considers that learning process is active construction of people's conceptual work, is a process which is based on inherent mental representation, is a process to obtain and construct new knowledge by interaction of old knowledge or experience with external environment. Humanism regards that learning can only rely on internal fact, not rely on teacher's external force. Learning activities should be selected and decided by students themselves; and in a good learning environment, students will learn everything. Cooperative learning theory is a teaching theory and strategy with creative thinking and actual effect. Collaborative learning is all learners' relevant activities of individual and group cooperation and mutual assistance of the learners who participate teams to achieve common learning goals with certain incentives mechanism, for obtaining maximum results and acquisition of individual and group objectives [9].The concept of learning theories can be formed in the way that can be introduced and help in creating of computer science and engineering course.

### 6.2 Strategy 2

For the approach to transform the education content in M-learning, there are many kind of methods. Educational content is delivered via many electronic media. For the learning style, E-learning can be divided into synchronous learning mode and asynchronous learning mode. A benefit of the synchronous (real time) M-learning is that a small bandwidth in which the materials of this mode are archived and stored so that they can be controlled at any time. In contrast, the synchronous M-learning mode has the benefit of communicating information without delay. To accomplish synchronous interaction interfacing with simple text, audio hide resources, whiteboard media and application sharing is needed. This mode also permits attendance checking by educators, allows floor control for question/answer and nonetheless it supports toolbox to achieve suitable learning interaction.

M-Learning is a new style of learning that comes from E-learning with the development of mobile communication technology and the popularity of the teams. M-Learning is suitable for the realization of E-learning easy to use mobile devices and wireless Internet. Therefore, it is more appropriate for the education of college students, and complements the E-Learning, which improves the flexibility, convenience, interactivity of E-Learning. The goal of M-learning is to achieve the 5 W (ie, any person, when, where, whoever, whatever) of learning, and M-learning is to create a learning community news, to promote the achievement of learning permanent. As in the eye of researchers in distance education, M-learning is considered the last step of distance education as follows: D-learning (distance learning), E-learning and M-learning [9] [10].

### 6.3 Strategy 3

We should notice when we design the educational information: on one hand, the information must be organized in sentences so that we can post it with cellular phones. On the other hand, learning contents design should interest students in discussing topics, so the topics should not be too difficult or too easy for students; them must be designed according to the students' cognitive level. The educators define the information and contents of the M-learning module. It comprises of a sequence of components that fabricate the complete effect of the module. However, they can also be evaluated individually. The series of elements define the course layout, the way it is communicated, the language used to communicate it and the information expressed. The course has to possess the following factors: content, elements, treatment, structure, and code.

### 6.4 Strategy 4

An interactive representation of M-learning is required in order to fabricate and maintain the learning resources. This interaction can occur among variety of groups such as student and student, student and educator, student and content, educator and content, content and content, and lastly educator and educator. This interaction is divided into two groups; one where the instructor assists in developing the design of the contents and activities of the M-learning model and second is where the instructor constantly organizes these contents and activities that have been constructed. The instructor may also update the contents if needed. Interactive model for M-learning systems are shown in Figure-2.

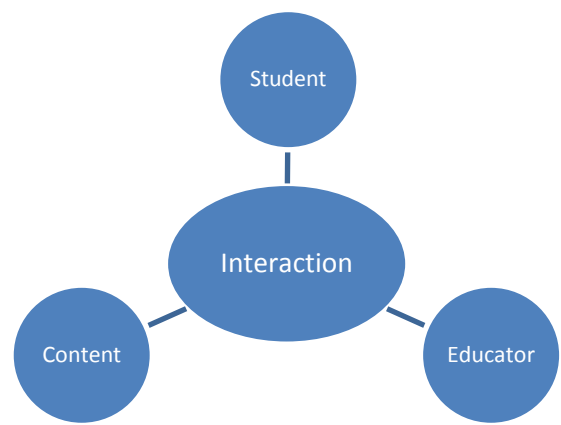

Figure 2: An interactive model of M-learning

### 6.4.1 Interaction Student-Student

The student-student interaction represents the interaction of peer-to-peer mobiles on M-learning. This environment offers the advantages of cognitive learning skills, allows one to gain social skills in education as well as improving personal bonds between participants and communities.

For engineering and computer science it's important to allow students to communicate between each other, this is needed in case of laboratory ,team project.

### 6.4.2 Interaction Student–Educator

With this interaction, a student is able to select their M-learning with instructor assistance, whereby a community of inquiry is created. This interaction module allows communication between students and their instructors to be netbased (synchronous or asynchronous) and could be carried out using different available formats including text, audio, video and so forth. These formats offer the benefit of maintaining large number of students' information or queries which have to be accounted for by the instructor or instantly responded to [11].

### 6.4.3 Interaction student–Content

The interaction between student and content is available in numerous formats, using mobile devices to access wireless websites, which offers learning contents. This module is known as independent study [11].

### 6.4.4 Interaction Content–Content

This module offers programs which automatically update the M-learning contents interacting with other contents. In doing so, mobile students' context will be efficiently supported.

### 6.4.5 Interaction Educator–Content

This module is beneficial in that it fabricates and maintains the learning resources. This interaction is divided into two groups; one whereby the instructors make the M-learning contents and activities and the other where the contents and activities that have been created are constantly refreshed and controlled [11].

### 6.4.6 Interaction Educator–Educator

This interaction module enhances the progress and support from an instructor through other instructors in the group. With the knowledge augmentation of the instructors in the community, the information swap among instructions can be increased [11].

## 6.5 Strategy 5:

Communication is the heart of the M-learning system for computer science and engineering core. When the sender transmits a message, it is a great necessity that the receiver recognizes this message. As a result, there has to be an agreement between the sender and receiver that the receiver is aware of the message sent, and fully understands it, in order to achieve a successful communication. Communication is accomplished through the use of a source (sender), a destination (receiver), a transmitter channel (medium) through which communication occurs. It is has been argued that communication can be affected by the following five attributes [12].

### 6.5.1 Immediacy of feedback

The significance of this characteristic is that it specifies the degree to which the medium allows the users to provide quick feedback regarding the information they receive. The function of the medium is to assist rapid bidirectional communication.

### 6.5.2 Parallelism

This feature corresponds to the number of participants. If the number of participants is small then parallelism is of unimportance. On the other hand, if the group of members participating is large, parallelism becomes a great convenience. Typically, convergence gains an advantage from low parallelism, because of the need to understand the viewpoints of each member. With the advantages that parallelism offers, it also consists of some drawbacks. One of these drawbacks is that noise is introduced due to the complexity of monitoring and coordinating the high number of separate conversations as well as developing the common understanding from them.

In case of a framework for integrating, individuals understanding voting structure are used to determine this. This is an exception and likely occurs in the production function than in the group well being or membership support function.

### 6.5.3 Feedback

Having the availability of feedback betters the system performance in such a way that it allows understanding between the communicating parties. This is achieved through mid-course corrections in the transmission of messages which further results in the correction of any false elements that may have been sent with a message. Thus, it can be concluded, having immediacy of feedback more or less composes of two major advantages; communication speed and accuracy. Having said that, with the advantages there are also some drawbacks which need to be address. The first problem faced is the cost. The sender and receiver must engage in a synchronous interaction. To do this, scheduling is required so that both agree on a time to interact; therefore significant efforts are needed in some circumstances. The second problem associated with feedback is that some media enable quick feedback, and in turn create high hopes for rapid feedback which can get in the way of communication. For instance, face to face communication needs fast feedback, which can interfere with deliberation, encouraging premature action. Thus, it can be concluded that for information which requires deliberation (e.g., large amount of information or complex information), rapid feedback may not be as beneficial and therefore can degrade the system performance. On the contrary, convergence needs less information and cognitive efforts than the deliberation, hence rapid feedback may not have as much of an affect as it does on the deliberation. Due to the idea that the aim is to understand interpretations of information as opposed to the information itself, feedback becomes compulsory.

### 6.5.4 Rehearsability

This feature allows one to fabricate a message consisting of the exact meaning that they intend. Rehearsability becomes extremely vital as the message approaches high complexity or equivocality as it helps better the understanding. On the other hand, the problem faced with high rehearsability media is the lower feedback that is typically come across.

### 6.5.5 Reprocessability

The Reprocessability aspect allows the receiver to continuously process the message to gain full understanding of the message received. Also, it has the advantage of allowing deliberation. As the volume, complexity and equivocality enhances in a message, this feature becomes for vital. Despite the information or communication procedure (whether conveyance or convergence) reprocessability results in improved understanding. Having said that, it plays a more important role to the conveyance communication process, as conveyance generally fabricates information that needs deliberation, and of course reprocessability is a necessity for deliberation.

Typically, where convergence is the main objective in group communication processes, media that provides high feedback and low parallelism is needed to achieve good performance.

However, when conveyance is the main objective, media that provides low feedback and high parallelism is needed to achieve good performance. Communication model for M-learning systems are shown in Figure-3.

**Communication processes:**

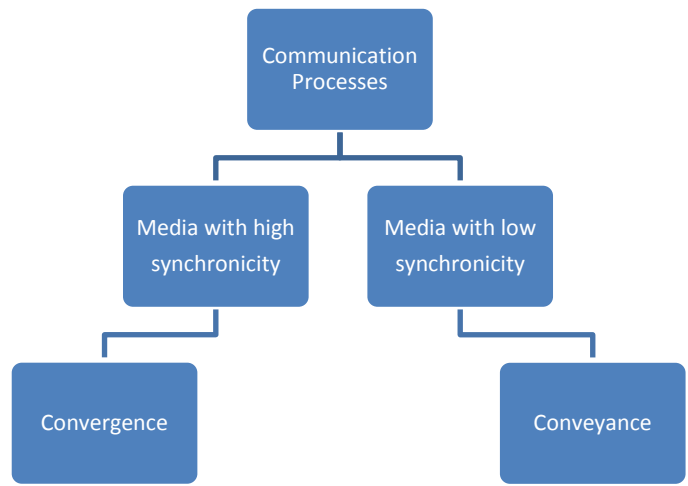

Figure 3: Group communication processes.

## 6.6 Strategy 6

There are probably several factors that influence the situation in which a given communication environment will be effective for different groups. One factor is the degree to which the groups worked together in the past. Established groups are more likely to have established standards for support group members and well-being (eg, roles within the group), and well-established rules for the processing of content. When the task is demand, it is likely that rules will only apply to the task. During the online class, members are able to work independently on assigned tasks. Online course requires more than just the convergence of transport, although some convergence is clearly needed. The need for media synchronicity is lower for the online class in creation.

However, when faced with a non-routine task that has little resemblance to the earlier work established groups may need more time creating and solving problems in the course material, and therefore require more convergence . As a group matures, they "may be able to perform all their needs, at least for projects, exchange of information with much less rich. This means that the communication needs of the groups may differ over time, the basis of shared experiences. 'perceptions about the usefulness of a means for a task and the group's ability to perform a task in an environment of change over time. The interaction of the greater experience of a group and the development of standards and standards, may interact and result in relatively improved performance of tasks in time. It is likely that group members get to know better over time, share common experiences that can be evoked by simple messages that relate to these experiences Shared.

## 6.7 Strategy 7

The integrated platform computing and engineering courses. Computer training and engineering (as currently organized) has the following components: [13]

- Lectures (in class).
- Homework assignments.
- Laboratory exercises.
- Midterm and final exams.

We have to create the structure of the courses that give us the prototype M-learning course content (of the class to convert a line).The course M-learning systems creator is shown in Figure-4.

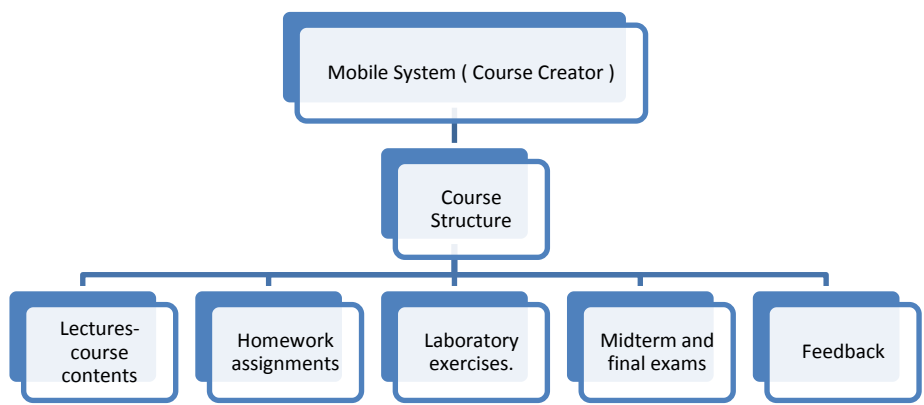

Figure 4: Course Creator

Students use smart phones with Wi-Fi functions could benefit from the ease of available hot water stain in the center of academic learning and many other hot spots in major cities. If students are invited to access the site and download the materials using a computer before transferring it to your mobile phone. Content developers should develop each M-learning material based on single mobile learning objectives using effective learning strategies and style, and formats. The use of text, graphics, audio, video and animation to support the desired learning outcomes. The content developer should strive to be innovative, while keeping everything short and simple.

## 7. CONCLUSION

M-learning can help provide a rich environment for teaching and learning tools for learning and developing high, and at the same time, improve the teaching skills of teachers, and promote the implementation of lifelong learning, lifelong learning and achieve the ultimate goal of training creative talents. Then M-learning strategies and guidelines are very important to be implemented in the development and the creation of M-learning application of course content. By implementing a user interface very good, it can help students gain more understanding of the topics covered in the application and a simple user interface will also reduce the time needed to consider the request. These strategies serve as a platform for the construction and connection to other services. This gave us the opportunity to begin to create new tools and services to enable students to achieve better learning outcomes.

As a result, teachers and trainers must design learning materials for delivery on a variety of mobile devices. Design of learning materials for mobile devices in accordance with good theories of learning and teaching of proper design and effective learning

## Authors

### Ibrahim M Alkore Alshalabi

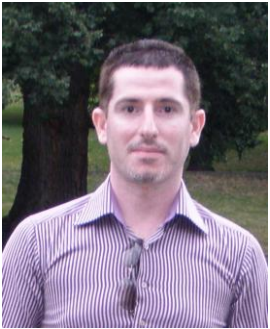

Ibrahim M Alkore Alshalabi received the B.Sc. in Computer Science from Al-Isra Private University, Amman, Jordan in 1997, and the MCA( Master of Computer Applications ) from Bangalore University - India in 2007. In 2009 he joined University of Bridgeport as Ph.D. student in computer science and engineering at the University of Bridgeport, Connecticut-USA. From 1997 to 2004, he was Assistant Lecturer in Ma'an Community College - Al-Balqa Applied University-Jordan. From 2007 to 2009 he joined Al-Hussein Bin Talal University-Jordan as assistant lecturer. Ibrahim M Alkore Alshalabi has research interest is in the general area of E-Learning, M-Learning, wireless communications and networks.He actively participated as a committee member of International Conference on Engineering Education, instructional technology, Assessment, and E-Learning (EIAE 10, EIAE 11).

### Khaled Elleithy

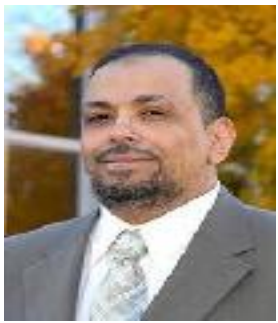

**Dr. Elleithy** is the Associate Dean for Graduate Studies in the School of Engineering at the University of Bridgeport. He has research interests are in the areas of network security, mobile communications, and formal approaches for design and verification. He has published more than one hundred fifty research papers in international journals and conferences in his areas of expertise. Dr. Elleithy is the co-chair of the International Joint Conferences on Computer, Information, and Systems Sciences, and Engineering (CISSE). CISSE is the first Engineering/Computing and Systems Research E-Conference in the world to be completely conducted online in real-time via the internet and was successfully running for four years.Dr. Elleithy is the editor or co-editor of 10 books published by Springer for advances on Innovations and Advanced Techniques in Systems, Computing Sciences and Software. Dr. Elleithy received the B.Sc. degree in computer science and automatic control from Alexandria University in 1983, the MS Degree in computer networks from the same university in 1986, and the MS and Ph.D. degrees in computer science from The Center for Advanced Computer Studies at the University of Louisiana at Lafayette in 1988 and 1990.